\documentclass[journal=jctcce, manuscript=article]{achemso}

\usepackage{chemformula} % Formula subscripts using \ch{}
\usepackage[T1]{fontenc} % Use modern font encodings
 \usepackage[colorlinks=true,linkcolor=blue,urlcolor=blue,citecolor=blue]{hyperref}
\usepackage{array}
\usepackage{lipsum}
\usepackage{bm}
\usepackage{subfigure}
\usepackage{amssymb}
\usepackage{multirow}
\usepackage{tabularx}
\usepackage{amsmath}
\usepackage{braket}
\usepackage{caption}
\usepackage{float}
\usepackage{caption}
\usepackage{enumerate}
\usepackage{ulem}
\usepackage{caption}

\usepackage{color}
\usepackage{mathtools}
\usepackage{graphicx}
\usepackage{dcolumn}
\usepackage{array}
\usepackage{lipsum}
\usepackage{bm}
\usepackage{subfigure}
\usepackage{amssymb}
\usepackage{multirow}
\usepackage{tabularx}
\usepackage{amsmath}
\usepackage{braket}
\usepackage{csquotes}
\graphicspath{{plots/}}
 \usepackage{lipsum}
\usepackage{mathrsfs}
\usepackage{MnSymbol}
\usepackage{dirtytalk}
\newcommand{\beq}{\begin{equation}}
\newcommand{\eeq}{\end{equation}}
\newcommand{\bea}{\begin{eqnarray}}
\newcommand{\eea}{\end{eqnarray}}

\renewcommand{\vec}[1]{\mathbf{#1}}

\SectionsOn
\SectionNumbersOn

\author{Zhandos Moldabekov}
\email{z.moldabekov@hzdr.de}
\affiliation{Center for Advanced Systems Understanding (CASUS), D-02826 G\"orlitz, Germany}
\alsoaffiliation{Helmholtz-Zentrum Dresden-Rossendorf (HZDR), D-01328 Dresden, Germany}

\author{Thomas D. Gawne}
\affiliation{Center for Advanced Systems Understanding (CASUS), D-02826 G\"orlitz, Germany}
\alsoaffiliation{Helmholtz-Zentrum Dresden-Rossendorf (HZDR), D-01328 Dresden, Germany}

\author{Sebastian Schwalbe}
\affiliation{Center for Advanced Systems Understanding (CASUS), D-02826 G\"orlitz, Germany}
\alsoaffiliation{Helmholtz-Zentrum Dresden-Rossendorf (HZDR), D-01328 Dresden, Germany}

\author{Thomas~R.~Preston}
\affiliation{European XFEL, D-22869 Schenefeld, Germany}

\author{Jan Vorberger}
\affiliation{Helmholtz-Zentrum Dresden-Rossendorf (HZDR), D-01328 Dresden, Germany}

\author{Tobias Dornheim}
\affiliation{Center for Advanced Systems Understanding (CASUS), D-02826 G\"orlitz, Germany}
\alsoaffiliation{Helmholtz-Zentrum Dresden-Rossendorf (HZDR), D-01328 Dresden, Germany}

\title{ Ultrafast Heating Induced Suppression of \textit{d}-band Dominance in the Electronic Excitation Spectrum of Cuprum}

\begin{document}

\begin{abstract}
The combination of isochoric heating of solids by free electron lasers (FEL) and \textit{in situ} diagnostics by X-ray Thomson scattering (XRTS) allows for measurements of material properties at warm dense matter (WDM) conditions relevant for astrophysics, inertial confinement fusion, and material science. In the case of metals, the FEL beam pumps energy directly into electrons with the lattice structure of ions being nearly unaffected. This leads to a unique transient state that gives rise to a set of interesting physical effects, which can serve as a reliable testing 
platform for WDM theories.
In this work, we present extensive linear-response time-dependent density functional theory (TDDFT) results for the electronic dynamic structure factor of isochorically heated copper with a face-centered cubic lattice. 
At ambient conditions, the plasmon
is heavily damped due to the presence
of $d$-band excitations, and its position is independent of the wavenumber.
In contrast,
the plasmon feature
starts to dominate the excitation spectrum and has a Bohm-Gross type plasmon dispersion for temperatures $T \geq 4~{\rm eV}$, where the quasi-free electrons in the interstitial region are in the WDM regime.
In addition, we analyze the thermal changes in the $d$-band excitations and outline the possibility to use future XRTS measurements of isochorically heated copper as a controlled testbed for WDM theories.
\end{abstract}

% \maketitle

\section{Introduction}

The study of matter under extreme densities and temperatures has emerged as a highly active research field due to the availability of modern laser facilities equipped with various X-ray diagnostic techniques.
High-power laser facilities are routinely being used to explore the physics and chemistry at conditions relevant to planetary astrophysics \cite{Millot2018, Pierleoni_PNAS_2010,Benuzzi_Mounaix_2014}, and inertial confinement fusion~\cite{hu_ICF, Fernandez2019PRL,Betti2023}, and to explore new exotic materials \cite{Miao2020Nature}. For example, using lasers for heating and compression allows one to measure macroscopic properties such as the equation of state ~\cite{Regan_PRL_2012,Falk_PRL_2014,Tilo_Nature_2023,Dornheim_T_2022,Dornheim_T2_2022,dornheim2024unraveling}.
In addition, ultrashort 
%free electron laser (FEL) beams at such 
X-ray free electron laser (XFEL) capabilities e.g.~at the European XFEL \cite{Zastrau_2021} and LCLS \cite{Fletcher2015} have opened the way to study phenomena on femtosecond timescales~\cite{Vorberger_PLA_2024}. By heating the electrons without directly affecting the ions, an XFEL with a sub-100 fs duration provides a unique opportunity to generate and study a transient state with hot electrons within the unperturbed crystal structure of the ions \cite{Descamps_sciadv, Descamps2020}. In these experiments, the X-ray Thomson scattering (XRTS) technique~\cite{siegfried_review} can then be used to probe the electronic structure of a given system by measuring its electronic dynamic structure factor (DSF) $S(\mathbf{q},\omega)$, where $\mathbf{q}$ and $\omega$ are the change in momentum and frequency of the scattered photon.
%properties since the XRTS spectrum is defined by the dynamic structure factor (DSF) of electrons.

%plying FEL laser, e.g., 
In this way,
it was shown that the laser-induced heating of electrons leads to the lattice instability and melting (disordering) of silicon due to the weakening of the inter-ionic bonds \cite{PhysRevB.26.1980, PhysRevLett.96.055503}.
% Another example is the laser-induced solid--solid phase transition that has been observed in bismuth \cite{Sokolowski-Tinten2003}.
In contrast to semiconductors, metals can remain stable under laser heating of the electrons, and, strikingly, can even manifest a more rigid lattice structure.
For example, Descamps et al.~\cite{Descamps_sciadv} have recently reported the observation of a stable gold crystal lattice where the electrons have been heated by the FEL to a few electronvolts.
In this experiment, a signature of phonon hardening has been observed, whereby the bonds stiffen between atoms. This effect was earlier predicted by \textit{ab initio} Kohn-Sham density functional theory (KS-DFT) calculations \cite{PRL_Recoules} 
of the hot electrons within the cold ionic lattice using the local density approximation (LDA) for the exchange--correlation (XC) functional.
%where KS-DFT with local density approximation for the exchange-correlation (XC) functional was used to compute the structure of hot electrons in a cold ionic lattice.
A second example for the successful utilization of KS-DFT for predicting the properties of solids with laser-excited electrons is the calculation of the XRTS spectrum of an isochorically heated aluminium
foil by Mo \textit{et al.}~\cite{Mo_2018} based on linear-response time-dependent DFT (LR-TDDFT) using an adiabatic LDA (ALDA) XC kernel.
%Another instance of an effective application of the KS-DFT method for the description of the experimental data for laser-irradiated solids is the linear-response KS-DFT (LR-TDDFT) simulation of the XRTS spectrum of isochorically heated aluminum foil by Mo et al \cite{Mo_2018}, where an adiabatic LDA (ALDA) XC kernel was used in the LR-TDDFT calculations. 
%The LR-TDDFT allows one to describe the XRTS measurements because the latter is defined by the DSF of electrons.
The same combination of LR-TDDFT with ALDA was shown to accurately describe the XRTS spectrum of aluminum at ambient conditions, where the plasmon was measured with ultrahigh resolution  at the European XFEL \cite{gawne_prl}.
Therefore, LR-TDDFT can be expected to yield accurate results for the electronic dynamic structure factor of isochorically heated metals across temperature regimes.

%Therefore, for metals for which KS-DFT provides accurate equilibrium state properties, the LR-TDDFT is expected to be a reliable tool for the description of the XRTS spectrum across temperature regimes. 

Very recently, Moldabekov \textit{et al.}~\cite{moldabekov2024excitation} have used this approach to study the effect of electronic heating on the order of a few electronvolts on the expected XRTS spectrum~\cite{moldabekov2024excitation};  this has revealed an interesting red shift of the plasmon energy by $0.1\,$eV for aluminium and by $1\,$eV for silicon as a consequence of thermal excitations.
%The results of the LR-TDDFT study of the effect of heating up to several electronvolts on the DSF of electrons in aluminum and silicon have shown a peculiar red shift of the plasmon energy in the limit of low wavenumbers due to thermal excitations \cite{moldabekov2024excitation}. 
%This red shift is shown to be around $0.1~{\rm eV}$ for aluminum and around $1~{\rm eV}$ for silicon. 
In the case of aluminium, the effect is small and only manifests at small wavenumbers $q\lesssim 0.1 {\rm \AA^{-1}}$, making it very challenging to measure.
%Due to its small value and emergence only at small wavenumbers $q\lesssim 0.1 {\rm \AA^{-1}}$, the heating-induced red shift in aluminum is highly challenging to measure. 
For silicon, on the other hand, the plasmon shift of $1~{\rm eV}$ at temperatures $T\simeq 2 {\rm eV}$ is well within experimental measurement capabilities \cite{Wollenweber_2021, gawne_prl}. However, the possible instability of the lattice due to the weakening of the inter-ionic bonds \cite{PhysRevB.26.1980, PhysRevLett.96.055503} can be a serious obstacle in practice.
Therefore, it is important to ask if such a heating-induced red shift prominently manifests itself in other metals that are stable under FEL radiation.  
Going back to aluminium, an additional thermally induced feature is the formation of a double plasmon peak as the region of Landau damping is shifted to lower wavenumbers upon increasing the electronic temperature~\cite{moldabekov2024excitation}.
%Another thermally induced feature in aluminum is the formation of the double-plasmon peak due to the shift of the Landau damping region to lower wavenumbers with the increase of the electronic temperature \cite{moldabekov2024excitation}.
This effect is similar to the formation of the double plasmon in the DSF of ground-state aluminum near the pair continuum  \cite{Schuelke, Larson_2000, Cazzaniga_2011}. These results show that thermal excitations in X-ray-driven solids can generate a variety of new features in the XRTS spectrum at a finite momentum transfer.

In the present work, we carry out extensive new LR-TDDFT calculations to explore the XRTS spectrum of isochorically heated copper.
In contrast to simple metals, the effect of \textit{d}-states dominates over plasmon-type excitations in the DSF of electrons in transition metals \cite{Alkauskas_2013, Rubio_PRB1999}.
In gold and copper, excitations originating in the \textit{d}-band lead to the formation of a prominent double peak structure at $\omega>\omega_p$ and a substantial broadening of the plasmon feature at $\omega=\omega_p$.  
Interestingly, the presence of the \textit{d}-state excitations leads to a plasmon dispersion that is nearly independent of the wavenumber for both materials \cite{Alkauskas_2013, Rubio_PRB1999}.
Here, we investigate in detail the interplay of these effects with thermal excitations on the DSF of copper at different temperatures, wavenumbers, and crystallographic directions.
Indeed, thermal effects on the DSF are profound: we find an emerging collective plasmon excitation that becomes dominant over the $d$-band feature for $T\gtrsim4\,$eV and which starts to follow the familiar Bohm-Gross relation in this regime. In addition, we find a pronounced blue-shift of the plasmon with increasing $T$, which is in stark contrast to other isochorically heated metals such as Al~\cite{moldabekov2024excitation}. Finally, we discuss the possibility to use XRTS experiments with isochorically heated copper as a rigorous testbed for the theoretical modeling of \emph{warm dense matter} (WDM)~\cite{wdm_book,new_POP,Dornheim_review}---an extreme state that occurs in astrophysical objects~\cite{Kritcher2020,Benuzzi_Mounaix_2014,becker,fortov_review}, and which plays an important role e.g.~for inertial confinement fusion~\cite{hu_ICF,Betti2016,Betti2023} applications.

%\textcolor{red}{TD: mention what we find in one or two sentences: 1) emergence of plasmon dispersion; 2) suppression of $d$-band dominance; 3) use Bohm-Gross fits [see Gawne~\cite{gawne2024ultrahigh}] to determine free electronic density and ionization $Z$ $\to$ isochoric heating particularly interesting case as mass density is known. New insights into models and stuff!!11}

%In this work, to get insight into the effect of the isochoric heating of electrons on the DSF dominated by excitations from \textit{d}-states in the ground state, we investigate in detail using the LR-TDDFT method 

The paper is organized as follows:
In Sec.~\ref{s:methods}, we give an overview of the LR-TDDFT approach and provide computational details of our simulations.
%We start by providing a summary of the related LR-TDDFT aspects and the details of the performed simulations in Sec. \ref{s:methods}. 
The results of the calculations are presented and discussed in Sec. \ref{s:results}. The paper is concluded by a summary of the main findings and an outlook over future works in Sec.~\ref{s:end}.

\section{LR-TDDFT approach to the dynamical structure factor}\label{s:methods}

\subsection{Theoretical Framework}

The intensity that is measured in an XRTS experiment is given by a convolution of  the combined source-and-instrument function $R(\omega_s)$~\cite{Dornheim_T2_2022} and the electronic dynamic structure factor $S(\mathbf{q},\omega)=S(\mathbf{q},-\Delta \omega)$ (with $\Delta \omega$ being the energy loss of the scattered photon),
\begin{eqnarray}
    I(\mathbf{q},\Delta \omega_s) = S(\mathbf{q},\Delta \omega) \circledast R(\omega_s) \ ,
\end{eqnarray}
where the latter accounts both for the finite width of the probing X-ray source, and for all effects of the detector~\cite{MacDonald_POP_2022}. The momentum transfer $\mathbf{q}$ is determined from the scattering angle.
%In experiments, the XRTS spectrum is obtained by probing a target using the X-ray laser beam of a finite width. Within the first Born approximation, the scattered signal from the target represents a convolution of the dynamical structure factor of electrons with the so-called source-and-instrument function representing the X-ray beam form and the used detector type. 
The state-of-the-art is given by the European XFEL in Germany, where XRTS measurements with the capability of resolving electronic features
%experiments are capable of resolving features
with a resolution of up to $\delta \omega\sim0.1\,{\rm eV}$ have been recently demonstrated \cite{gawne_prl}.

To study the effect of thermal electronic excitations on the XRTS spectrum of X-ray-driven copper, we use the LR-TDDFT method with an adiabatic XC kernel.
Indeed, LR-TDDFT constitutes the most common method to study the DSF of solids, and there is a vast body of dedicated literature, see e.g.Ref.~\cite{book_Ullrich} and references therein; here, we restrict ourselves to a concise overview of the main ideas.

%Here we provide a brief description of the LR-TDDFT theory used for the DSF calculations in this work. 
As a first step, we consider the well-known fluctuation--dissipation theorem that connects the macroscopic dielectric function $\epsilon_M(\vec q, \omega)$ with $S(\mathbf{q},\omega)$~\cite{Kubo_1966, quantum_theory},
\begin{equation}\label{eq:DSF}
    S(\vec q, \omega)=-\frac{\hbar^2q^2}{4\pi^2e^2n} \frac{1}{1-e^{-\hbar \omega/k_BT}} {\rm Im}\left[\epsilon_M^{-1}(\vec q, \omega)\right],
\end{equation}
where $n$ denotes the electronic number density, and $e$ the elementary charge.
The term  ``{macroscopic}'' indicates that $\epsilon_M(\vec q, \omega)$ describes the volume averaged response to an external perturbation \cite{Gurtubay_PRB_2005, Moldabekov_PRR_2023, JCP_averaging}.  
It is computed by taking the diagonal part of the inverse microscopic dielectric matrix $\epsilon_M^{-1}(\vec q, \omega)= \left[\epsilon^{-1}(\vec k, \omega)\right]_{\vec G\vec G}$, where $\vec q=\vec G+ \vec k$ (with  $\vec k$ being in the first Brillouin zone) and $\vec G$ is a reciprocal lattice vector \cite{DSF_LR-TDDFT, martin_reining_ceperley_2016}. The latter is defined by the microscopic density response function ~\cite{book_Ullrich},
\begin{equation}\label{eq:d_f}
      \varepsilon^{-1}_{\scriptscriptstyle \vec G,\vec G^{\prime}}(\vec k,\omega)=\delta_{\scriptscriptstyle \vec G,\vec G^{\prime}}+\frac{4\pi}{\left|\vec k+\vec G\right|^{2}}  \chi_{\scriptscriptstyle \vec G,\vec G^{\prime}} (\vec k,\omega).
\end{equation}

The LR-TDDFT method allows one to compute $ \chi_{\scriptscriptstyle \vec G,\vec G^{\prime}} (\vec k,\omega)$ in different approximations. The lowest rank corresponds to the so-called independent particle approximation (IPA).
In the IPA, the Kohn-Sham (KS) orbitals and eigenenergies are used to calculate the density response function $\chi^{~0}_{\scriptscriptstyle \vec G,\vec G^{\prime}}(\vec k,\omega)$ according to the ideal electron gas model \cite{Hybertsen}. 
Since the KS eigenenergies from the self-consistent ground state (equilibrium state) calculations are employed, $\chi^{~0}_{\scriptscriptstyle \vec G,\vec G^{\prime}}(\vec k,\omega)$ already has information about excitations between different orbitals. However, being computed using a formula for the ideal Fermi gas model,  $\chi^{~0}_{\scriptscriptstyle \vec G,\vec G^{\prime}}(\vec k,\omega)$ omits various correlation effects, such as screening due to the Hartree mean field, microscopic density inhomogeneities due to the field of the ions etc.  The inclusion of correlation effects leads to a Dyson’s type equation for the density response function  $ \chi_{\scriptscriptstyle \vec G,\vec G^{\prime}} (\vec k,\omega)$  \cite{Byun_2020, martin_reining_ceperley_2016}:
\begin{equation}\label{eq:Dyson}
\begin{split}
\chi_{\scriptscriptstyle \mathbf G \mathbf G^{\prime}}(\mathbf k, \omega)
&= \chi^0_{\scriptscriptstyle \mathbf G \mathbf G^{\prime}}(\mathbf k, \omega)+ \displaystyle\smashoperator{\sum_{\scriptscriptstyle \mathbf G_1 \mathbf G_2}} \chi^0_{\scriptscriptstyle \mathbf G \mathbf G_1}(\mathbf k, \omega) \big[ v_{\scriptscriptstyle \mathbf G1}(\vec k)\delta_{\scriptscriptstyle \mathbf G_1 \mathbf G_2} \\
&+ K^{\rm xc}_{\scriptscriptstyle \mathbf G_1 \mathbf G_2}(\mathbf k, \omega) \big]\chi_{\scriptscriptstyle \mathbf G_2 \mathbf G^{\prime}}(\mathbf k, \omega),
\end{split}
\end{equation}
where $v_{\scriptscriptstyle \mathbf G1}(\vec k)={4\pi}/{|\mathbf k+\mathbf G_1|^2}$ is the Coulomb potential in reciprocal-space, and $ K^{\rm xc}_{\scriptscriptstyle \vec G_1,\vec G_2}(\vec k, \omega)$ is the XC kernel capturing electronic correlations; it is defined as the functional derivative of the XC potential in KS-DFT \cite{Gross_PRL1985}.

The LR-TDDFT method provides the DSF of the electrons in the thermal equilibrium.
An alternative approach that can perform the simulation of electronic dynamics with a distribution different from the Fermi-Dirac distribution is real-time TDDFT (RT-TDDFT), where electronic wave functions are propagated according to time-dependent KS equations. This method was used by Silaeva \textit{et al.} \cite{Silaeva_prb_2018} to study ultrafast electron dynamics thermalization in metals driven by a 7-fs laser pulse. Silaeva \textit{et al.}\cite{Silaeva_prb_2018} showed that valence electrons reach a thermalized state within the time of the laser pulse. The RT-TDDFT method can also be used to compute the DSF. For example, Baczewski \textit{et al.}~\cite{Baczewski_prl_2016} used RT-TDDFT to compute the DSF of warm dense beryllium in thermal equilibrium. 
We note that if the same XC functional were used in both, RT-TDDFT and LR-TDDFT are formally equivalent for linear response properties in thermal equilibrium ~\cite{book_Ullrich}.

In our calculations, we have used a static (adiabatic) XC kernel   $K^{\rm xc}_{\scriptscriptstyle \vec G_1,\vec G_2}(\vec k, \omega=0)$ within the ALDA~\cite{book_Ullrich}.
The ALDA is known to provide a fairly accurate description of the macroscopic dielectric function $\epsilon_M(\vec q, \omega)$ of metals and semiconductors at finite wavenumbers \cite{gawne_prl, Quong_PRL, Weissker_PRL, Cazzaniga_2011, Si_Weissker}.  The relevant thermal signatures explored in this work are characterized by a difference of $\delta \omega \gtrsim 1~{\rm eV}$ from the ground state features. 
If needed, a further fine-tuning can be achieved either by employing more advanced static XC kernels beyond ALDA \cite{Moldabekov_non_empirical_hybrid, Moldabekov_JCTC_2023} or by using an explicitly dynamic approximation (e.g., see Ref.~\cite{Byun_2020} and references therein) in future works.
%However, we feel that such computationally costly investigations should be postponed until experimental measurements have become available.

%However, being highly computationally expensive, such fine-tuning is better to be done if the experimental measurements show the ALDA to be insufficient at relevant parameters. 

We note that, in the ground state, the DSF is usually studied indirectly by measuring the electronic energy loss spectrum (EELS), see, e.g., Refs.~\cite{Krane_JPF_1978,sprosser1989aluminum,brydson2020electron}.
In principle, the XRTS spectrum and the EELS are directly related since
\begin{equation}
    {\rm EELS}(\vec q, \omega)=-{\rm Im}\left[\epsilon_M^{-1}(\vec q, \omega)\right].
\end{equation}

From Eq.~(\ref{eq:DSF}), one can see that $S(\vec q, \omega)\sim q^2~{\rm EELS}(\vec q, \omega)$. This means that EELS is advantageous for measurements at small wavenumbers, whereas XRTS might be more suitable at large wavenumbers.
However, EELS measurements are problematic for experiments with matter under extreme conditions due to its requirements for thin targets as well as long measurement times~\cite{brydson2020electron}, which are not realistic for the transient states that are of interest in the current work.

In consistency with measurements at FEL facilities, we consider the electronic response on subpicosecond time scales and treat the ions as being frozen in their crystal lattice positions surrounded by heated electrons.
This is justified since the electron-lattice equilibration time is order of picoseconds \cite{Cho_SciRep_2016, Cho_prl_2011, Jourdain_prl_2021, Smirnov_prb_2020, Nicoul_AplPhysLett_2011, White_prb_2014}.  Furthermore, this approximation is corroborated by the predictions of increased melting temperature with electron heating in copper and other d-band metals \cite{Recoules_prl_2006, Smirnov_prb_2020}, and by the recent observation of phonon hardening in gold \cite{Descamps_sciadv}.

\subsection{Calculation parameters}

We used the GPAW code~\cite{GPAW1, GPAW2, LRT_GPAW1, LRT_GPAW2, ase-paper, ase-paper2}, which is a real-space implementation of the projector augmented-wave (PAW) approach~\cite{BlochlPAW}. We used the ground-state LDA XC-functional by Perdew and Wang \cite{Perdew_Wang}.
The simulations have been carried out for a face-centered cubic (fcc) lattice with the lattice parameter $3.61$~\AA~set according to the experimental value~\cite{wyckoff1948crystal}.
For the calculation of the KS states, we used the energy cutoff $1000~{\rm eV}$, the PAW dataset of copper provided by GPAW (with $1s-3p$ orbitals treated as frozen core electrons), and the primitive cell combined with
the $k$-point grid $40\times40\times40$. We note that in the employed LR-TDDFT formalism, the momentum transfer must be the difference between two $k$-points times $2\pi/a$.
To study the possible impact of inhomogeneity with respect to crystallographic directions on the DSF, calculations were performed along the [100], [111], and [011] directions.
We considered electronic temperatures in the range $0.025~{\rm eV}\leq T \leq 12~\rm{eV}$ and the number of KS bands was set to $N_b=100$. The smearing of the occupation numbers was computed according to the Fermi-Dirac distribution.
On the stage of the calculation of the density response matrix, the local field effect cutoff was set to $150~{\rm eV}$. In all calculations, we used $\eta=0.1~{\rm eV}$ for the Lorentzian smearing parameter in $ \chi_{\scriptscriptstyle \vec G,\vec G^{\prime}} (\vec k,\omega)$ \cite{LRT_GPAW2}.  For the density of state (DOS) calculations of copper, we used the same parameters as for the DSF.  The DOS was plotted by setting the Gaussian width parameter to $0.2~{\rm eV}$.

\section{Simulations Results and Discussion} \label{s:results}
\subsection{DSF $S(\vec q, \omega)$ in the X-ray driven copper}\label{ss:DSF}

\begin{figure*}[t!]
    \centering
    \includegraphics[width=0.75\textwidth,keepaspectratio]{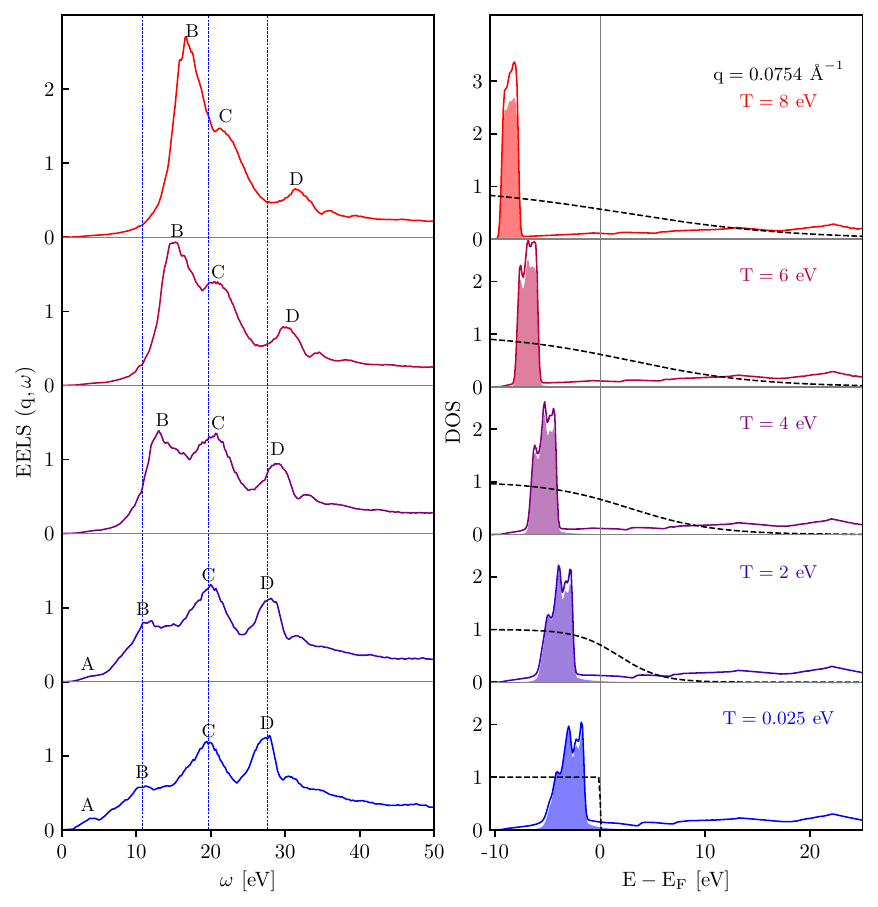}
    \caption{Left panel: EELS spectrum along the [100] direction. Right panel: total DOS (solid lines), projected DOS on $d$-orbital (shaded), and Fermi-Dirac occupation number distribution (dashed lines).
    Shown are results for $q=0.0754$ \AA\textsuperscript{-1} at ambient conditions [$T=0.025~{\rm eV}$], at $T=2~{\rm eV}$, $T=4~{\rm eV}$, $T=6~{\rm eV}$, and at $T=8~{\rm eV}$.}
    \label{fig:EELS}
\end{figure*}
\begin{figure}[t!]
    \centering
    \includegraphics[width=0.6\textwidth,keepaspectratio]{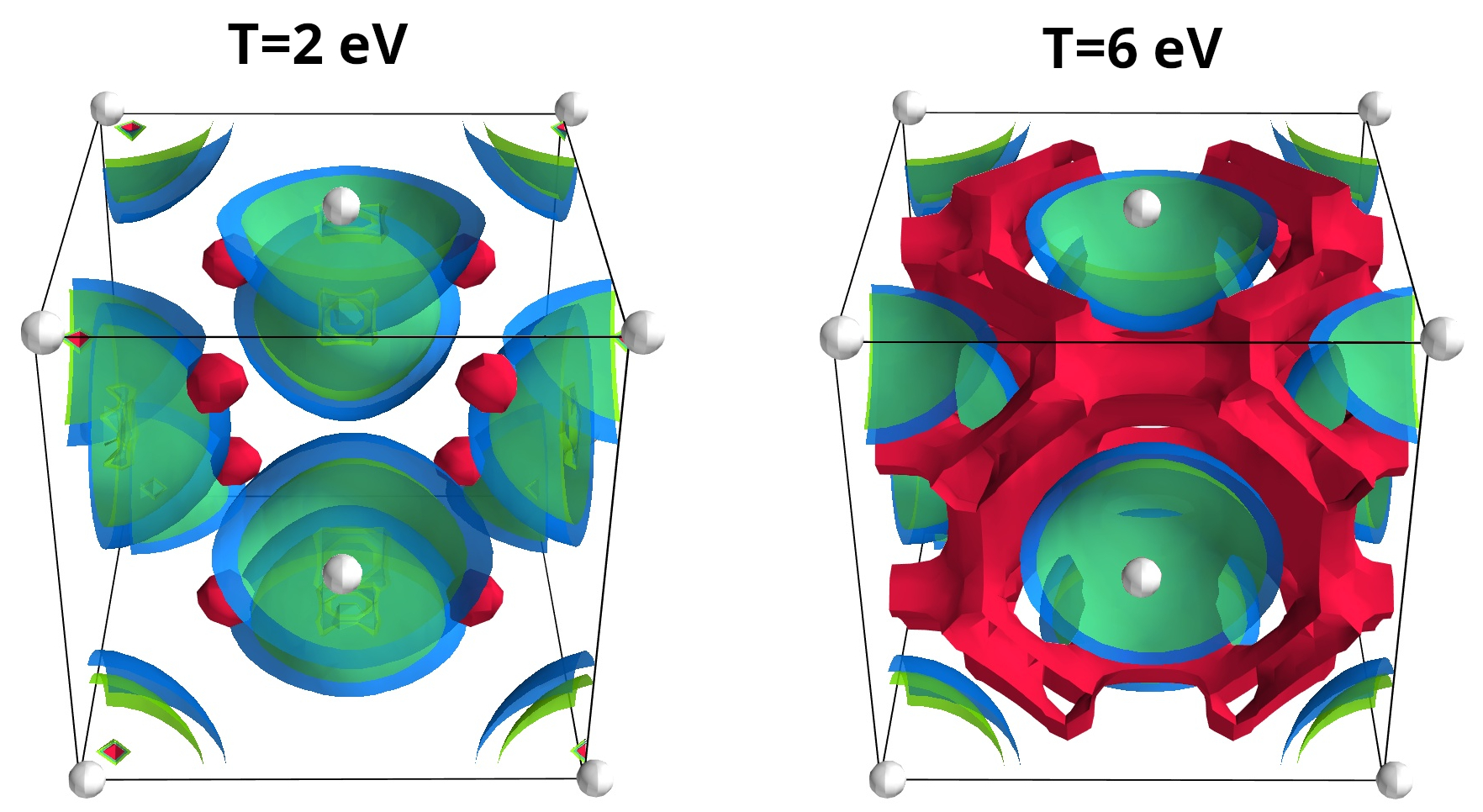}
    \caption{Electronic density accumulation in the interstitial region and density depletion around the ions
   % Illustration of the interstitial regions with the electron density accumulation and depletion 
   due to heating at $T=2~{\rm eV}$ (left) and $T=6~{\rm eV}$ (right). The surface plots (semi-transparent) indicate the density change with respect to the ground state, $\delta n(\vec r)=n_T(\vec r)-n_0(\vec r)$, i.e., relative to the density at $T=0.025~{\rm eV}$. The blue surface corresponds to $\delta n(\vec r)=0$, the red surface indicates $\delta n(\vec r)={\rm max}[\delta n(\vec r)]/70>0$, and the green surface indicates $\delta n(\vec r)={\rm min}[\delta n(\vec r)]/15<0$.} 
    \label{fig:3Dsurf}
\end{figure}

\begin{figure*}[t!]
    \centering
    \includegraphics[width=1\textwidth,keepaspectratio]{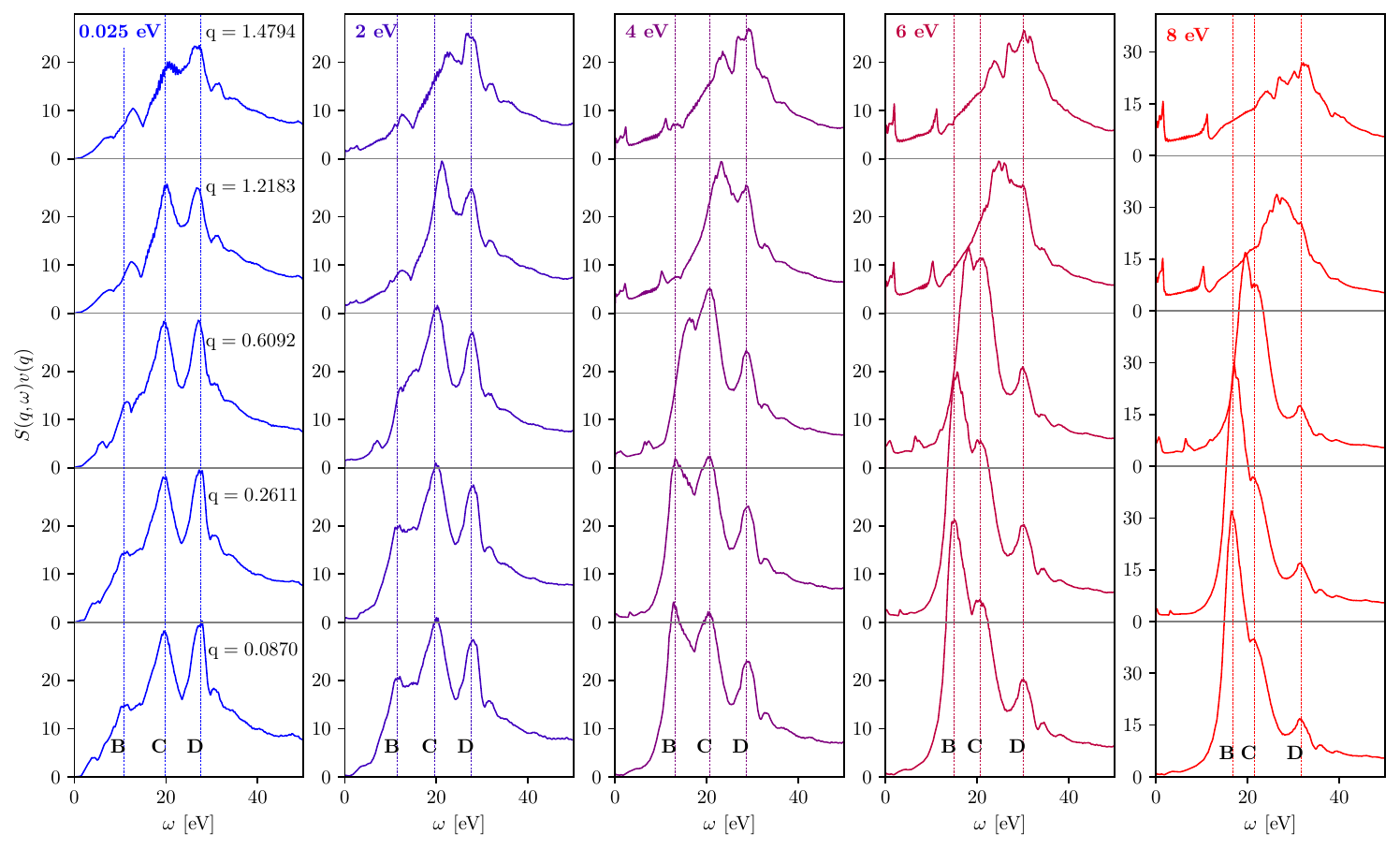}
    \caption{LR-TDDFT results for the DSF of fcc Cu along the [011] direction at different wavenumbers for the ground state with $T=0.025~{\rm eV}$, and for heated electrons with $T=2~{\rm eV}$,  $T=4~{\rm eV}$, $T=6~{\rm eV}$, and $T=8~{\rm eV}$. The wavenumber values are given in the units of \AA\textsuperscript{-1}.} 
    \label{fig:DSF011}
\end{figure*}

We start our investigation by considering the electronic ground state (here represented by the results for $T=0.025\,$eV) in the limit of small wave numbers.
In the left panel of Fig.~\ref{fig:EELS}, we show the EELS spectrum at 
%
%We start by considering the results in the limit of small wavenumbers. 
%Since previous low-temperature calculations were focused on EELS properties \cite{Alkauskas_2013, Rubio_PRB1999}, we first look at the EELS results for the ground state (here represented by the results for $T=0.025~{\rm eV}$) and consider the influence of heating on the EELS.
%In the left panel of Fig. \ref{fig:EELS}, we show the EELS at
$q=0.0754$ \AA\textsuperscript{-1} for different temperatures in the range from 0.025 eV up to 8 eV (with the temperature increasing in the subplots from bottom to top).
We focus on four main features of the EELS spectrum (which equivalently appear in the DSF, cf.~Fig.~\ref{fig:DSF011}) denoted by capital letters A--D. % in the left panel of Fig. \ref{fig:EELS}.
A thorough investigation of the EELS properties of copper in the ground state has been presented by Alkauskas et al \cite{Alkauskas_2013}, where it was shown that features A and B are plasmon type collective oscillations, whereas 
C and D are a consequence of  excitations between the \textit{d}-band and the unoccupied states above the Fermi level.
More specifically, peak B can be described as a collective plasmon oscillations of the valence electrons; this has been shown by Campillo et al \cite{Rubio_PRB1999} by freezing the \textit{d}-band into the core and leaving only the $4s^1$ state that forms the valence electron. Our LR-TDDFT results are in good agreement with the structure of electronic excitations reported in Refs. \cite{Alkauskas_2013, Rubio_PRB1999}, and we reproduce the positions of all four peaks A--D.

Let us next turn to the central topic of this work, which is the study of the impact of thermal effects.
With increasing the temperature,
the positions of peaks B, C, and D shift to larger energies. 
Moreover, the amplitude of the plasmon feature B gets substantially amplified.
In contrast, the signature of feature A nearly vanishes for $T\geq 2~{\rm eV}$. % due to thermal excitations. 
Compared to peaks A and B, the magnitude of features C and D are only weakly affected by heating.

Following the analysis by Campillo et al \cite{Rubio_PRB1999} who investigated the DSF of copper at ambient conditions,
signatures B and C can be understood by considering the density of states (DOS), which is shown in the right panel of Fig.~\ref{fig:EELS} as the solid lines.
%In the right panel of Fig. \ref{fig:EELS}, we show the DOS of electrons at different temperatures (solid lines). 
The $ d$-band dominates the accumulation of the states below the Fermi energy.
This is shown by the projected density of states on $d$-states, which is depicted by the shaded area.
We find that, with increasing temperatures, the \textit{d}-states are shifted to lower energies.
% This is a consequence of the reduced occupation number 
% \textcolor{red}{which leads to a reduced electronic screening and, therefore, a lower binding energy [cite]}
% , see also the corresponding Fermi-Dirac distributions that are shown as the dashed black lines in the same panel.
Since the features C and D in the EELS/DSF emerge as the result of the transitions from \textit{d}-states to high-lying bands above the Fermi level, the observed shift of the \textit{d}-states 
results in a blue-shift % features C and D 
by approximately the same amount.
An accurate determination of the difference in the shifts of the C and D peaks is difficult due to the strong broadening of the C peak.  Nevertheless, one can observe that the blue-shift of C and D peaks have close values at $T<4~{\rm eV}$. We estimate that the D peak of the EELS shifts by about $1 ~{\rm eV}$ more towards larger frequencies at $T=8~{\rm eV}$ and $T=6~{\rm eV}$ compared to the C peak.

%The shift of the  $d$-state energy is due to the reduction of the occupation numbers of the $d$-states. 
%This can be understood by looking at the Fermi-Dirac distribution (dashed lines) provided in the right panel of Fig. \ref{fig:EELS}. 

%The latter results in an increase in the frequency of the plasmon (feature B in the left panel of Fig \ref{fig:EELS}).

In addition, increasing the temperature leads to an increased number of conduction electrons, which, in turn, leads to a larger plasmon frequency, i.e., a substantial blue-shift of feature B.
The corresponding increase in the electronic density in the interstitial region between the fixed ions can be demonstrated by investigating the change in the electronic density $\delta n(\vec r)$ with respect to the ground state.
For $T=2~{\rm eV}$ ($T=6~{\rm eV}$), in atomic units, we find ${\rm max}[\delta n(\vec r)]\simeq1.45$ (${\rm max}[\delta n(\vec r)]\simeq3.974$) and ${\rm min}[\delta n(\vec r)]\simeq-0.375$ (${\rm min}[\delta n(\vec r)]\simeq-0.966$).
This is illustrated in more detail in Fig.~\ref{fig:3Dsurf}, where we show the thermally induced density change in the 3D simulation box for both temperatures.
%For these temperatures, in Fig. \ref{fig:3Dsurf}, we illustrate the change in density due to heating. 
The blue surface depicts $\delta n(\vec r)=0$; the red surface indicates $\delta n(\vec r)>0$ (for the illustration, we choose $\delta n(\vec r)={\rm max}[\delta n(\vec r)]/70$); 
the green surface indicates $\delta n(\vec r)<0$ (here we show $\delta n(\vec r)={\rm min}[\delta n(\vec r)]/15$). 
We can thus clearly see that the electronic density is reduced in the vicinity of the ions and accumulates in the interstitial region, which is more pronounced for the higher temperature, as it is expected.
A more detailed, quantitative analysis of the increased free electronic density and the resulting plasmon blue-shift is presented in Sec.~\ref{ss:plasmon} below.

\begin{figure*}[t!]
    \centering
    \includegraphics[width=1\textwidth,keepaspectratio]{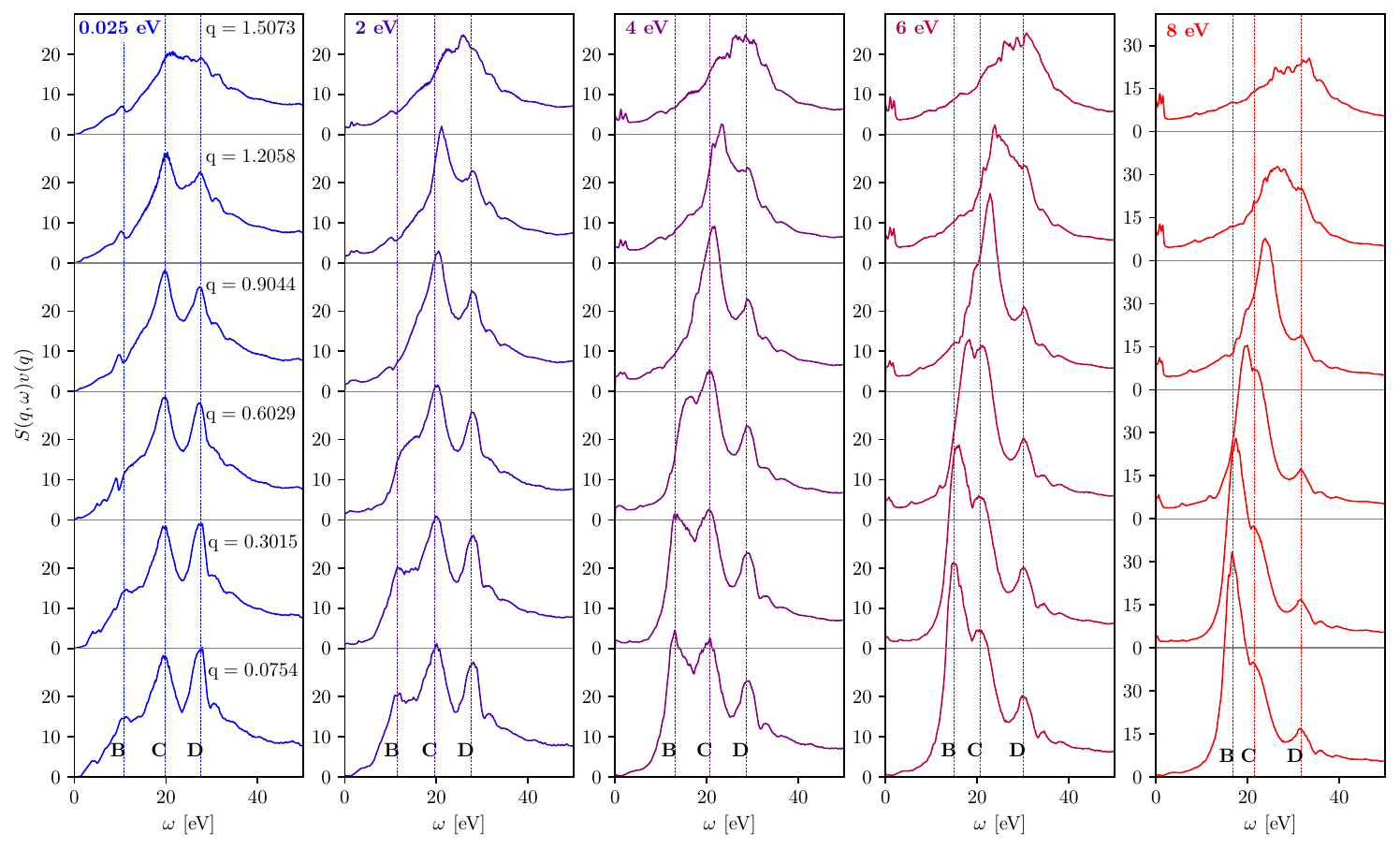}
    \caption{LR-TDDFT results for the DSF of fcc Cu along the [100] direction at different wavenumbers for the ground state with $T=0.025~{\rm eV}$, and for isochorically heated electrons with $T=2~{\rm eV}$,  $T=4~{\rm eV}$, $T=6~{\rm eV}$, and $T=8~{\rm eV}$. The wavenumber values are given in the units of \AA\textsuperscript{-1}.} 
    \label{fig:DSF100}
\end{figure*}

Let us next investigate the DSF, which is the key property in XRTS experiments;
%Next, we analyze the change in the DSF with the increase in wavenumber at different temperatures. 
it is shown in Fig. \ref{fig:DSF100} 
%, we provide the results for the DSF in 
along the [100] direction for $0.0754$~\AA\textsuperscript{-1}$\leq q \leq 1.4794$~\AA\textsuperscript{-1}.
Since feature A is strongly damped for $T\geq 2~{\rm eV}$, we will ignore it in the following discussion and instead focus on the thermally induced changes of features B, C, and D.
At all considered temperatures, the position of features C and D is nearly independent of the wavenumber. 
In stark contrast, the plasmon feature B exhibits a considerably richer behavior.
At $T=0.025~{\rm eV}$ and at $T=2~{\rm eV}$, its position does not follow the Bohm-Gross type dispersion of the free electron gas~\cite{quantum_theory}. Overall, it is difficult to quantify its $q$-dependence for these two temperatures due to the comparably weak spectral weight and the possible overlap with other features due to local field effects created by the lattice structure \cite{Alkauskas_2013}.
In contrast, we observe a pronounced increase with $q$  for $T\geq 4 {\rm eV}$.
As a result, the plasmon eventually overtakes feature C, leading to the disappearance of the latter from the DSF at $q\gtrsim 0.9$~\AA\textsuperscript{-1} in the cases of  $T=4~{\rm eV}$ and $T=6~{\rm eV}$, and at $q\gtrsim 0.6$ \AA\textsuperscript{-1} for $T=8~{\rm eV}$. 

An additional interesting research topic is due to the lattice structure, which is known to lead to an anisotropy of the DSF with respect to the crystallographic direction at certain wavenumbers~\cite{PhysRevB.40.5799, Alkauskas_2013}. 
To quantify this anisotropy effect on X-ray driven copper, we show the DSF in the [011] direction at different temperatures and wavenumbers in Fig. \ref{fig:DSF011}. We find that the DSF in the [011] direction closely resembles the DSF in the [100] direction for $q\lesssim 0.9$~\AA\textsuperscript{-1}, and differences in the shape and peak positions emerge for $q\gtrsim 1.2$~\AA\textsuperscript{-1}.
For completeness, we note that the DSF in [111] direction is equivalent to the [100] direction (see Appendix).
In summary, we conclude that the electronic DSF of copper is nearly isotropic at $T\geq 2~{\rm eV}$ for $q\lesssim 0.9$~\AA\textsuperscript{-1}.

\begin{figure}[t!]
    \centering
    \includegraphics[width=0.5\textwidth,keepaspectratio]{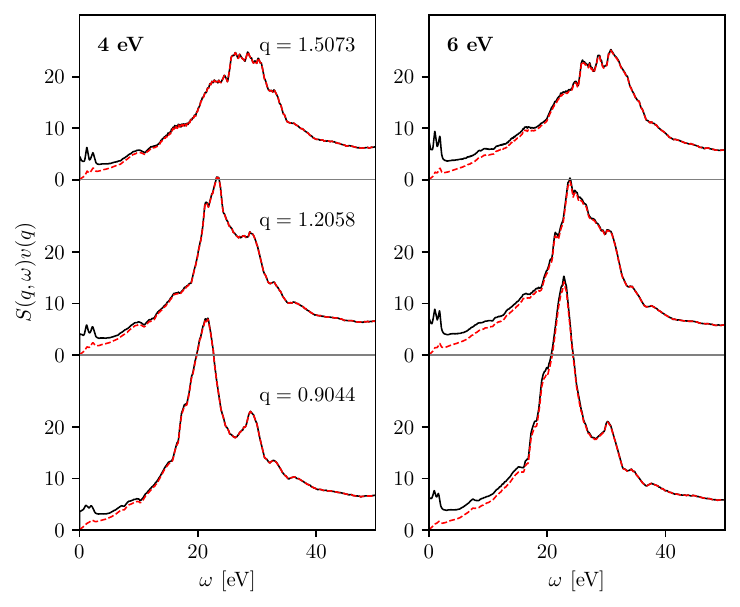}
    \caption{The demonstration of the role of the factor $f(\omega)=\left(1- e^{-\hbar \omega/k_BT}\right)^{-1}$ on the enhancement of the DSF features at small energies.
    Solid lines are the results for the DSF along the [100] direction and dashed lines are the same data divided by $f(\omega)$.   The wavenumber values are given in the units of \AA\textsuperscript{-1}. } 
    \label{fig:DSFvsEELS}
\end{figure}

In addition to the discussed dominant features, we observe an increase in the DSF at low energies and a new peaked feature emerges in the DSF at $\omega<10~{\rm eV}$  in both [100] and [011] directions at $T\geq 4~{\rm eV}$ and $q\gtrsim 0.6$~\AA\textsuperscript{-1}. 
The increase in the temperature modifies the DSF due to the factor $f(\omega)=\left(1- e^{-\hbar \omega/k_BT} \right)^{-1}$ in Eq. (\ref{eq:DSF}). This effect is particularly pronounced for the low-energy part of the DSF, and enhances subtle features in this regime. This is illustrated in Fig. \ref{fig:DSFvsEELS} for $T=4~{\rm eV}$ and $T=6~{\rm eV}$ at wavenumbers $q= 0.9044$~\AA\textsuperscript{-1}, $q= 1.2058$~\AA\textsuperscript{-1}, and $q= 1.5073$~\AA\textsuperscript{-1}. From Fig. \ref{fig:DSFvsEELS} we observe that the factor $f(\omega)$ in Eq. (\ref{eq:DSF}) for the DSF results in the substantial increase of the DSF values at low energies leading to an amplification of the thermally induced features at $\omega<10~{\rm eV}$.
Physically, these features might originate from transitions between accumulated states located closely above the Fermi level, cf.~Fig.~\ref{fig:EELS}. Indeed,  
%from the left [right, no? :)] panel of Fig. \ref{fig:EELS}, 
one can observe that at $T\geq 4~{\rm eV}$ states above the Fermi level become partially occupied allowing the emergence of new excitation features.

\subsection{Plasmon dispersion and conditions in interstitial regions}\label{ss:plasmon}

\begin{figure}
    \centering
    \includegraphics[width=0.5\textwidth,keepaspectratio]{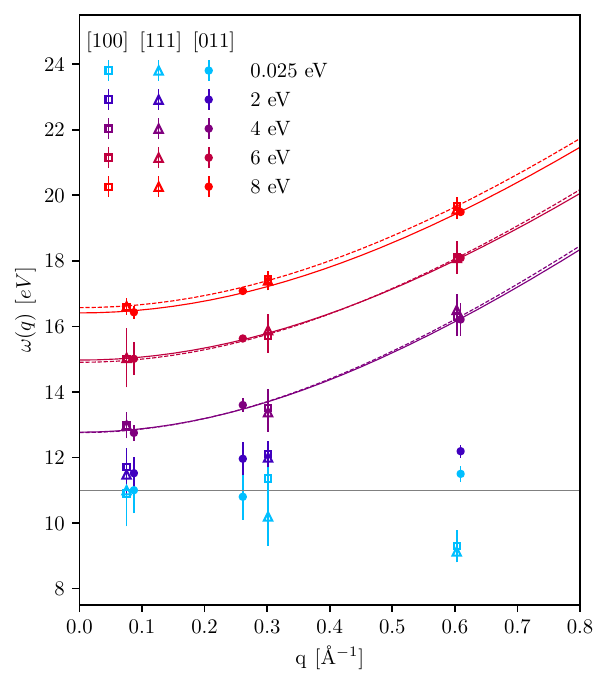}
    \caption{Dependence of the plasmon frequency on the wavenumber along different crystallographic directions at $T=0.025~{\rm eV}$, $T=2~{\rm eV}$,  $T=4~{\rm eV}$, $T=6~{\rm eV}$, and $T=8~{\rm eV}$. Solid (dashed) lines show Bohm-Gross type quadratic dispersion fits [cf.~Eq.~(\ref{eq:BohmGross})] for the direction [111] ([100]) at $T=4~{\rm eV}$, $T=6~{\rm eV}$, and $T=8~{\rm eV}$.
    In Bohm-Gross dispersion (\ref{eq:BohmGross}) we used for [111] ([100]) $\alpha=1.66\,\omega_p^2$ ($\alpha=1.7\,\omega_p^2$) at $T=4~{\rm eV}$,  $\alpha=1.24\,\omega_p^2$ ($\alpha=1.3\,\omega_p^2$) at $T=6~{\rm eV}$, and  $\alpha=1.11\,\omega_p^2$ ( $\alpha=1.25\,\omega_p^2$) at $T=8~{\rm eV}$.
    The horizontal solid grey line at $10.864~{\rm eV}$ indicates the plasmon energy computed using the free electron gas model, i.e., assigning the $3d$ shell as a core state [e.g. see Ref. \cite{Rubio_PRB1999}]. 
    %%%%%%%%%% Not true: \textcolor{red}{assuming a single valence electron for the effective density}
    } 
    \label{fig:wp011}
\end{figure}

%From the analysis of the DSF, it follows 
Analyzing the DSF, we have found
that the collective plasmon oscillations in X-ray driven copper overcome the dominance of the $d$-band excitations, eventually overtaking them with respect to the spectral weight. % due to the coupling of $d$-states with the states above the Fermi level.  
%This is evident from the amplitude of the signature of the plasmon compared to the signatures of the excitations from the $d$-states. 
In addition, %it is indicated by the fact that 
the plasmon position starts to exhibit a substantial dispersion with respect to the wavenumber for sufficiently high temperatures, which is in contrast to the ground state plasmon.
To examine the character of the plasmon dispersion, we show the dependence of the plasmon energy (frequency) on the wavenumber, $\omega(q)$, in Fig.~\ref{fig:wp011} for different temperatures. We consider  $q< 1$~\AA\textsuperscript{-1}, where the plasmon peak can be clearly identified at all considered temperatures, and independent of the crystallographic direction.
The uncertainty in the plasmon position is evaluated by looking at the onset of a broadened peak.
At $T=2~{\rm eV}$, we find that $\omega(q)$ is qualitatively similar to the results for $T=0.025~{\rm eV}$.
For $T\geq 4~{\rm eV}$, the plasmon dispersion starts to follow the familiar quadratic dependence on $q$ that is well-known from the free electron gas model~\cite{mahan1990many}.
To further quantify this trend, we fit the LR-TDDFT results using the Bohm-Gross type dispersion relation \cite{Bohm_Gross, Hamann_cpp},
\begin{equation}
    \omega^2(q)=\omega_p^2+\alpha q^2, 
    \label{eq:BohmGross}
\end{equation}
where $\alpha$ and $\omega_p$ are the free parameters. %The plasma frequency $\omega_p$ is inferred from the results at $q< 0.1 ~{\rm \AA^{-1}}$.
The results are shown as the solid ([011] direction) and dashed ([100] direction) lines in Fig.~\ref{fig:wp011}, which are nearly identical; 
%These two sets of curves are nearly identical with
the small differences are likely due to uncertainties introduced by the broadened peaks of the DSF.

In subsection \ref{ss:DSF}, %being a manifestation of the quasi-free electrons, we showed that the increase in the plasmon energy is due to an increase in the electron density in the interstitial region.
we have indicated that the increase in the plasmon energy with the temperature is a consequence of the excess electronic density in the interstitial regions between the ionic lattice, cf.~Fig.~\ref{fig:3Dsurf}.
Here, we propose to utilize such forward scattering data for the DSF as a diagnostic for the free electronic density and for the effective charge state.
%Therefore, the data for $\omega_p$ can be used to inquire parameters of the electrons in the interstitial space. 
Specifically, we define the effective density parameter $\widetilde{r}_s$ by inverting the usual relation between the density and the plasmon frequency of a free electron gas,
\begin{eqnarray}\label{eq:free_density_parameter}
  \omega_p = \left(4\pi\widetilde n \frac{e^2}{m_e}\right)^{1/2}\quad \Rightarrow  \quad \widetilde{r}_s = \left( \frac{3}{4\pi}\widetilde n \right)^{1/3}\ .
\end{eqnarray}
%$\widetilde{r_s}=(3/(4\pi \widetilde n))^{1/3}$
%of the quasi-free electrons by inverting the equation for the plasma frequency of free electron gas $\omega_p=(4\pi \widetilde n e^2/m)^{1/2}$.
We note that the parameter $\widetilde r_s$ also characterizes the coupling strength or, equivalently, the degree of non-ideality of the free electrons \cite{new_POP,Ott2018}.

An additional, related effective parameter of interest is given by the corresponding degeneracy temperature $\widetilde\Theta=k_BT/E_F(\widetilde n)$, where $E_F(\widetilde n)=(3\pi^2 \widetilde n)^{1/3}$ denotes the Fermi energy of a free electron gas of density $\widetilde n$~\cite{quantum_theory}. 
%At a given temperature $T$, the effective degeneracy parameter of the electrons in the interstitial region can be evaluated using $\widetilde\Theta=k_BT/\widetilde E_F$, where $\widetilde E_F=(3\pi^2 \widetilde n)^{1/3}$.
Finally, we consider the effective ionic charge $\widetilde Z$,
% \textcolor{red}{Please give formula!}, 
which is a key ingredient to equation-of-state tables~\cite{Falk_PRL_2014,Tilo_Nature_2023}.
The effective charge is computed using $\widetilde Z=\left(\widetilde r_s(T=0)/\widetilde r_s(T)\right)^3$, which follows from the condition $\widetilde Z n_i= \widetilde n$, with $n_i={\rm const}$ being the number density of ions, and setting $Z=1$ for copper at $T=0$ since we have one valence electron in the 4s state.

In Table \ref{t:wp_rs}, we provide an overview of these parameters for all selected temperatures.
%The results for $\omega_p$ from the LR-TDDFT simulations as well as  corresponding $\widetilde{r_s}$ and $\widetilde\Theta$ parameters at $0.025~{\rm eV}\leq T \leq 12~{\rm eV}$ are  provided in Table \ref{t:wp_rs}. 
For $T=0.025~{\rm eV}$, we find an effective density parameter of $\widetilde r_s=2.64\pm0.11$, which agrees with the value 
$r_s=2.668$ that corresponds to the conduction electrons density in the fcc copper at room temperature.
% ; the same holds for the effective plasma frequency $\omega_p=11\pm0.7\,$eV and the corresponding theoretical value of $\omega_p=10.815\,$eV. 

%We note that the $\widetilde r_s=2.64\pm0.11$ at $T=0.025~{\rm eV}$ has a close value to the density parameter $r_s=2.668$ computed counting only $1s$ electrons in fcc copper; with the corresponding free electron gas plasma frequency $\omega_p=10.815$.

%In addition, using information about plasmon frequency and the charge neutrality condition in the long-wavelength limit, one can evaluate the effective charge of ions $\widetilde Z$, which is an important concept for various warm dense matter and dense plasma models \textcolor{red}{[cite]}.
%We also provide evaluated $\widetilde Z$ values in Table \ref{t:wp_rs}.

Upon increasing the temperature, we find a monotonic increase in the density of effectively free electrons, leading to a corresponding decrease of $\widetilde{r}_s$ and an increase in $\widetilde Z$. The observed monotonic increase of $\widetilde\Theta$ is less trivial. On the one hand, we have $\widetilde\Theta\sim\widetilde{r}_s^2$~\cite{Ott2018}, which, by itself, would indicate a decrease of $\widetilde\Theta$ with $T$. However, this effect is overridden by the relation $\widetilde\Theta\sim T$ in practice.

Table \ref{t:wp_rs} clearly shows that the effectively free electrons in the interstitial region of isochorically heated copper are in the WDM regime~\cite{wdm_book}.
In nature, WDM occurs in a variety of astrophysical objects such as giant planet interiors~\cite{Benuzzi_Mounaix_2014}, brown dwarfs~\cite{becker}, and white dwarf atmospheres~\cite{Kritcher2020}. In the laboratory, these conditions are encountered on the compression path of a fuel capsule and its ablator in inertial confinement fusion experiments~\cite{hu_ICF,Betti2016,Betti2023}, and, in addition, they are relevant for material science, synthesis, and discovery~\cite{Kraus2016,Lazicki2021}.
Despite its fundamental importance for a gamut of
applications, the rigorous theoretical description of WDM remains challenging: it must cover the complex interplay between effects such as partial ionization, Coulomb coupling, and quantum degeneracy and diffraction, which is notoriously difficult in practice~\cite{wdm_book,new_POP,Dornheim_review}.
Our new simulation results thus imply that future XRTS experiments with isochorically heated copper constitute a suitable and highly controlled testbed for the benchmarking of theoretical methods and simulations.

%it follows that the electrons in the interstitial region of the X-ray-driven copper are in the warm dense matter regime (WDM) \cite{wdm_book}.
%The WDM is characterized by the correlated electrons in a partially degenerate state, which exists in the interiors of planets, and stars, and is generated in the ICF experiments.
%Therefore,  the data from XRTS experiments using an isochorically heated copper can be used as benchmark for WDM theories.
%This is particularly the case for large wavenumbers, where the effect of the XC kernel is strong \cite{Bohme_PRL_2022, Moldabekov_JCTC_2023}. 

%%%%%%%%%%%%%%%%%%%%%%%%%%%%%%%%%%%%%%%%%%%%% 
 \begin{table}[h!]
    \caption{Plasmon energy at $q=0.0754$~\AA\textsuperscript{-1} extracted from the DSF calculations with corresponding effective density parameter $\widetilde r_s$, effective degeneracy parameter $\widetilde \Theta$, and effective charge of ions $\widetilde Z$.\vspace*{0.3cm} }
    \label{t:wp_rs}
    \centering
\begin{tabular}{ |c|c|c|c|c|  }
  \hline
 T [eV] & $\omega_p$ [eV] & $\widetilde{r_s}$ & $\widetilde{\Theta}$& $\widetilde{Z}$\\
 \hline
  \hline
 0.025  & $11\pm 0.7$    &  $2.64\pm0.11$ &   $0$&   $1.0\pm0.06$\\
 2 & $11.52\pm 0.5$    & $2.56\pm0.07$   & $0.261\pm0.01$&   $1.1\pm0.09$\\
 4 & $12.75\pm 0.25$ & $2.39\pm 0.03$ &  $0.456\pm0.01$&   $1.34\pm 0.14$\\
 6  & $15.01\pm 0.5$ & $2.14\pm0.04$ &  $0.551\pm0.02$&   $1.86\pm0.17$\\
 8 &  $16.43\pm0.2$   & $2.02\pm0.02$ &$0.651\pm 0.01$&   $2.23\pm0.25$\\
 10 & $18.3\pm0.4$  & $1.88\pm0.027$   &$0.704\pm0.02$&   $2.77\pm0.29$\\
 12 & $19.43\pm 0.35$  & $1.80\pm0.02$ &$0.780\pm0.019$&   $3.12\pm0.33$\\
 \hline
\end{tabular}
\end{table}
%%%%%%%%%%%%%%%%%%%%%%%%%%%%%%%%%%%

%%%%%%%%%%%%%%%%%%%%%%%%%%%%%%%%%%%%%%%%%%%%%%%%%

\section{Conclusions}\label{s:end}

%In this work, m
Motivated by experimental capabilities at modern FEL facilities, we have performed a detailed study of the changes in the electronic DSF in fcc copper due to isochoric heating.
Our new LR-TDDFT simulations show that the heating induces a prominent plasmon feature that eventually becomes dominant over the $d$-band-signal.
This is in marked contrast to the ground state, where the plasmon is strongly damped by the presence of $d$-band excitations, and where it does not exhibit
%due to the excitations from the $d$-states to the empty states above the Fermi energy and does not show
a meaningful dispersion with respect to the wavenumber $q$.
Indeed, we have shown that at $T\geq 4~{\rm eV}$, the plasmon dispersion follows the familiar Bohm-Gross type relation, and the plasma frequency $\omega_p$ substantially increases with the temperature.
This has been explained by the accumulation of effectively free electrons in the interstitial region between the ions.
%due to thermal excitations as well as the accumulation of the electrons in the interstitial region.
This, in turn, is a consequence of the availability of the electrons in the $d$-orbitals for filling states in the quasicontinuum when the temperature is increased.
Interestingly, the reported behavior of the plasmon in copper is fundamentally different from isochorically heated aluminium (Al), where LR-TDDFT simulations have shown that heating to a few electronvolts (at $T\lesssim 7~{\rm eV}$) causes a red shift of  $\omega_p$~\cite{moldabekov2024excitation}. In this regard, we note that Al has 3 valence electrons from 3p3s orbitals and further ionization of the electrons from 2p requires temperatures about $70~{\rm eV}$. Therefore, in contrast to copper, thermal excitations in Al are not accompanied by an increase in the density of valence electrons with the increase in the temperature at $T \lesssim 10 ~{\rm eV}$.

From a physical perspective, we find that the quasi-free electrons in the interstitial space are in the WDM regime with an effective density parameter $\widetilde r_s\sim 2$ and an effective degeneracy temperature $\widetilde \Theta\sim 0.5$.
In addition to being interesting in their own right, such extreme conditions occur in a wealth of astrophysical objects and play a key role in experiments with inertial confinement fusion and material science.
In practice, there does not exist a single method that is capable of giving an accurate description of WDM states over the entire relevant parameter space, and the interpretation, modeling, and design of corresponding experiments is usually based on a number of de-facto uncontrolled approximations.

In this regard, we propose to use future XRTS experiments with isochorically heated copper as a controlled testbed for the rigorous assessment of different theoretical models and simulation tools for the description of WDM.
Ideally, one might infer the temperature of the heated sample based on the model-free imaginary-time thermometry approach~\cite{Dornheim_T_2022,Dornheim_T2_2022} as a first step. Indeed, even the inference of comparably moderate temperatures of $T\sim1-10\,$eV is expected to be feasible using the new high-resolution set-up that has recently been demonstrated at the European XFEL in Germany~\cite{gawne_prl}.
Second, we propose to carry out measurements at multiple scattering angles (i.e., multiple wavenumbers $q$) to observe the plasmon dispersion $\omega(q)$ and, in this way, to infer the effective charge state $\widetilde Z$ and the effective density parameter $\widetilde{r}_s$. Since the ambient density is a-priori known, this will give one full access to the most relevant parameters of the system.
This is an important advantage over shock experiments, or XRTS measurements with backlighter sources~\cite{MacDonald_POP_2022}, where one or multiple of these parameters can only be inferred on the basis of the theoretical models which we aim to test in the first place.
Isochoric heating can be achieved by employing a 400 nm optical short-pulse laser on thin targets and combining it with the delayed FEL probe \cite{Mo_Science_2018, Chen_prl_2013}. 
Alternatively, one can use X-ray pump and X-ray probe beams that are separated in both colour and time. Proper time separation allows heating to be completed by the pump before the target is probed. Such X-ray pump X-ray probe experiments can be performed at European XFEL~\cite{Liu2023} and at SACLA in Japan~\cite{hara2013two,inoue2020two}. A detailed discussion of the feasibility of measuring thermal excitations in the DSF of isochorically heated targets is provided in Ref.~\cite{moldabekov2024excitation}.

Such a hypothetical XRTS dataset can then be used to benchmark the zoo of available theoretical methods such as the widely used effective chemical models~\cite{siegfried_review,kraus_xrts,boehme2023evidence}.
A particularly interesting question is the assessment of XC functionals in KS-DFT simulations, including the resolution of explicitly thermal XC effects~\cite{thermal_PBE_2023, karasiev_gga_18,ksdt,groth_prl,review,karasiev_importance}.
Moreover, frequency-resolved inelastic X-ray scattering data for a set of finite wavenumbers $q$ will be ideally suited to gauge the accuracy of different
XC-kernels in LR-TDDFT calculations, including a rigorous assessment of the popular adiabatic approximation~\cite{book_Ullrich,Moldabekov_PRR_2023,dornheim_dynamic}. As a WDM testbed, the observed isotropy of the DSF of electrons in copper is advantageous for achieving unambiguous conclusions in the assessment of various models.

\appendix

\section*{Acknowledgments}
This work was partially supported by the Center for Advanced Systems Understanding (CASUS), financed by Germany’s Federal Ministry of Education and Research (BMBF) and the Saxon state government out of the State budget approved by the Saxon State Parliament.
This work has received funding from the European Union's Just Transition Fund (JTF) within the project \emph{R\"ontgenlaser-Optimierung der Laserfusion} (ROLF), contract number 5086999001, co-financed by the Saxon state government out of the State budget approved by the Saxon State Parliament.
This work has received funding from the European Research Council (ERC) under the European Union’s Horizon 2022 research and innovation programme
(Grant agreement No. 101076233, "PREXTREME"). 
Views and opinions expressed are however those of the authors only and do not necessarily reflect those of the European Union or the European Research Council Executive Agency. Neither the European Union nor the granting authority can be held responsible for them. Computations were performed on a Bull Cluster at the Center for Information Services and High-Performance Computing (ZIH) at Technische Universit\"at Dresden, at the Norddeutscher Verbund f\"ur Hoch- und H\"ochstleistungsrechnen (HLRN) under grant mvp00024, and on the HoreKa supercomputer funded by the Ministry of Science, Research and the Arts Baden-W\"urttemberg and
by the Federal Ministry of Education and Research.
% \end{acknowledgments}

\section*{Appendix}\label{s:app}

\begin{figure*}
    \centering
    \includegraphics[width=1\textwidth,keepaspectratio]{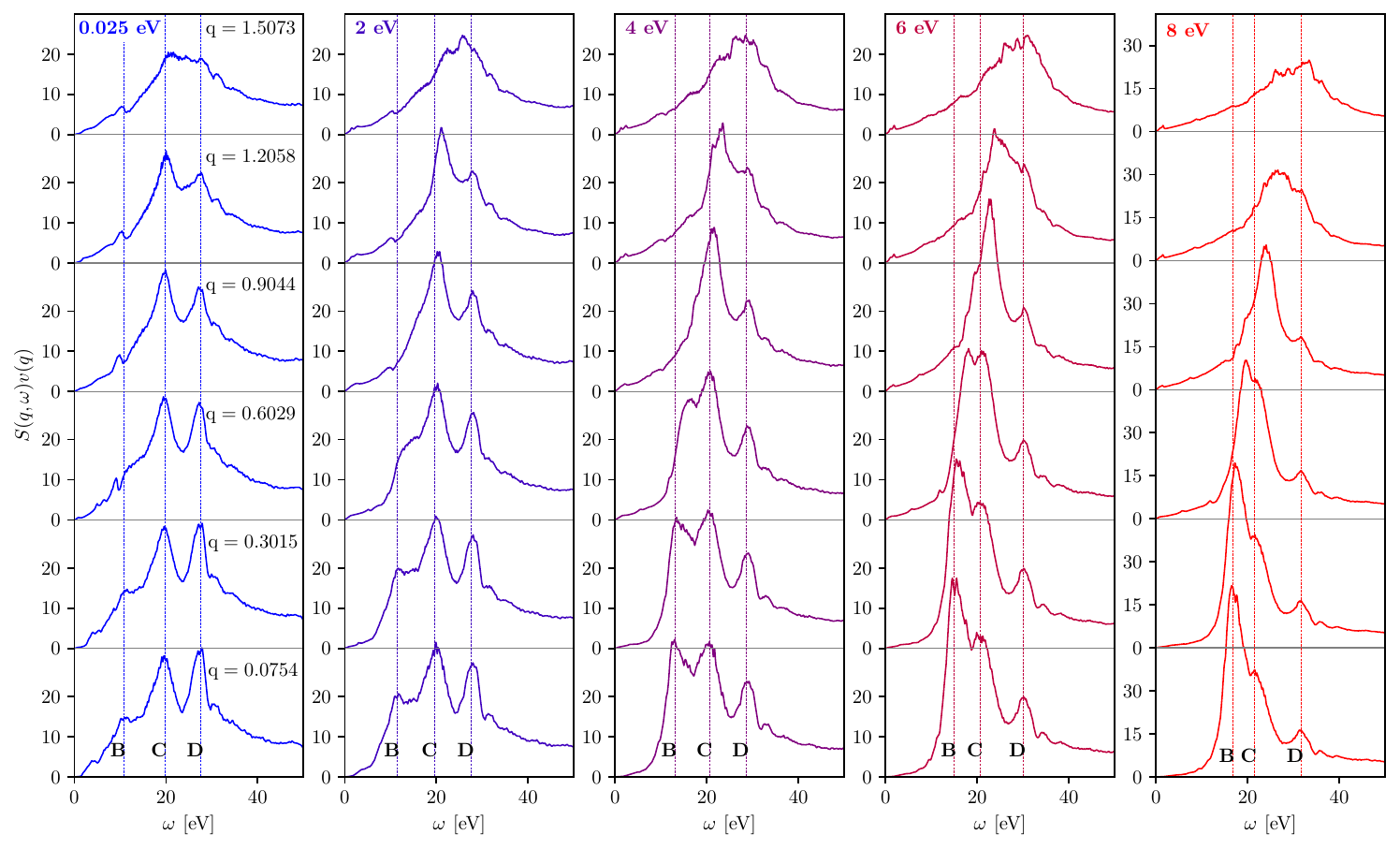}
    \caption{LR-TDDFT results for the DSF of fcc Cu along the [111] direction at different wavenumbers for the ground state with $T=0.025~{\rm eV}$, and for heated electrons with $T=2~{\rm eV}$,  $T=4~{\rm eV}$, $T=6~{\rm eV}$, and $T=8~{\rm eV}$. The wavenumber values are given in the units of~\AA\textsuperscript{-1}.} 
    \label{fig:DSF111}
\end{figure*}

For completeness, we show the DSF in the [111] direction at the different considered temperatures and wavenumbers in Fig.~\ref{fig:DSF111}.
We find that the B, C, and D features behave similarly to our results for the [100] and [111] directions, both with respect to temperature and wavenumber.
Indeed, the DSF in [111] direction is nearly identical to the DSF in [100] direction over the entire depicted $q$-range.

%%%%%%%%%%%%%%%%%%%%%%%%%%%%%%%%%%%%%%%%%%%%%%%%%%%%%%%%%%%%%%%%%%%%%%%%%%%%%%%%
% literature
%%%%%%%%%%%%%%%%%%%%%%%%%%%%%%%%%%%%%%%%%%%%%%%%%%%%%%%%%%%%%%%%%%%%%%%%%%%%%%%%
\bibliography{bibliography}
\end{document}